\documentclass[12pt,authoryear]{elsarticle}

\usepackage{graphicx}
\usepackage{amssymb}
\usepackage{url}
\usepackage{setspace}
\usepackage{lineno}
\usepackage{float}

\journal{Planetary and Space Science}
\begin{document}
\begin{frontmatter}

\title{Electron Dynamics near Diamagnetic Regions of Comet 67P/Churyumov-Gerasimenko}

\author[2]{H. Madanian\corref{cor1}}
\author[2]{J. L. Burch}
\author[3]{A. I. Eriksson}
\author[4]{T. E. Cravens}
\author[5]{M. Galand}
\author[3]{E. Vigren}
\author[2]{R. Goldstein}
\author[6]{Z. Nemeth}
\author[2]{P. Mokashi}
\author[7]{I. Richter}
\author[8]{M. Rubin}

\address[2]{Southwest Research Institute, 6220 Culebra Road, San Antonio, TX 78238, United Sates}
\cortext[cor1]{hmadanian@swri.edu}
\address[3]{Swedish Institute of Space Physics, Uppsala, Sweden}
\address[4]{Department of Physics and Astronomy, University of Kansas, Lawrence, KS, USA}
\address[5]{Department of Physics, Imperial College London, Prince Consort Road, London, UK}
\address[6]{Wigner Research Centre for Physics, Budapest, Hungary}
\address[7]{Institut für Geophysik und extraterrestrische Physik, TU Braunschweig, Braunschweig, Germany}
\address[8]{Physikalisches Institut, University of Bern, Bern, Switzerland}

\begin{abstract}
The Rosetta spacecraft detected transient and sporadic diamagnetic regions around comet 67P/Churyumov-Gerasimenko. In this paper we present a statistical analysis of bulk and suprathermal electron dynamics, as well as a case study of suprathermal electron pitch angle distributions (PADs) near a diamagnetic region. Bulk electron densities are correlated with the local neutral density and we find a distinct enhancement in electron densities measured over the southern latitudes of the comet. Flux of suprathermal electrons with energies between tens of eV to a couple of hundred eV decreases each time the spacecraft enters a diamagnetic region. We propose a mechanism in which this reduction can be explained by solar wind electrons that are tied to the magnetic field and after having been transported adiabatically in a decaying magnetic field environment, have limited access to the diamagnetic regions. Our analysis shows that suprathermal electron PADs evolve from an almost isotropic outside the diamagnetic cavity to a field-aligned distribution near the boundary. Electron transport becomes chaotic and non-adiabatic when electron gyroradius becomes comparable to the size of the magnetic field line curvature, which determines the upper energy limit of the flux variation. This study is based on Rosetta observations at around 200 km cometocentric distance when the comet was at 1.24 AU from the Sun and during the southern summer cometary season. 
\end{abstract}
\end{frontmatter}
\section{Introduction}
A cometary atmosphere is formed through heating and sublimation of ice and other volatiles at the comet surface. This mixture of neutral particles expands radially outward and is exposed to solar photons, solar wind charged particles, and the interplanetary magnetic field (IMF) \citep{glassmeier_interaction_2017,gombosi_physics_2015,cravens_cometary_2004}. As comets approach perihelion, sublimation and outgassing rates increase and plasma boundaries can be formed in the thicker coma against the impinging solar wind \citep{mandt_rpc_2016}. A bow shock is the outer most boundary that may be formed around comets where, due to assimilation of cometary ions into the solar wind, the supersonic flow slows down to subsonic speeds. Within the bow shock the solar wind plasma becomes heated. Upstream of comet Halley\textsc{\char13}s bow shock at about $10^6$ km from the nucleus, field-aligned backstreaming electrons were observed which were reflected from the enhanced magnetic field regions at the shock \citep{larson_electron_1992}. In addition, some perpendicularly heated electrons that escaped the cometary magnetosphere and traveled upstream along the magnetic field were present in the distribution at around $90^{\circ}$ pitch angles. \citet{fuselier_heat_1986} used data from the International Cometary Explorer (ICE) spacecraft at comet Giacobini-Zinner and showed similar distributions and increased electron heat flux backstreaming from the region of enhanced magnetic field near the comet.

At closer distances to the nucleus, a boundary that is particularly important for nonmagnetized objects is the diamagnetic cavity boundary. The magnetometer on the Giotto spacecraft detected the diamagnetic boundary around comet Halley at a cometocentric distance of 4500 km during the flyby inbound and the spacecraft exited the cavity at about 4150 km outbound \citep{neubauer_first_1986}. This global and relatively symmetric diamagnetic cavity is formed when the outward ion-neutral drag force in the cometary atmosphere balances the magnetic pressure gradient in the pile up region \citep{cravens_physics_1986,ip_formation_1987}. The size of the diamagnetic cavity at comet Halley was much larger than the solar wind proton gyroradius and magnetohydrodynamic models sufficiently described the stand-off distance of the diamagnetic boundary and the magnetic field profile across it \citep{lindgren_magnetohydrodynamic_1997,rubin_plasma_2014}. The AMPTE (Active Magnetospheric Particle Tracer Explorers) artificial comet experiment in 1984 created a temporary diamagnetic cavity ($\sim$60 s long and 70 km in radius) by releasing 2 kg of Barium vapor in the solar wind \citep{bingham_theory_1991,haerendel_dynamics_1986,gurnett_waves_1986}. The cavity boundary in this case was formed by electron currents. As \citet{haerendel_dynamics_1986} describe, photoelectrons of the expanding barium gas are coupled to ions via a polarization electric field, which further accelerates the ions radially outward. The electron gas initially reaches a pressure balance with the solar wind magnetic field while ions continue to expand, resulting in an inward polarization electric field. Under these conditions, electrons form a current layer as they undergo \textbf{E}$\times$\textbf{B} drift, leading to a shielding diamagnetic boundary.

At comet 67P/Churyumov-Gerasimenko (or 67P for short), similar diamagnetic regions have been observed, though the formation mechanism for these events is not yet fully understood. The magnetometer system onboard the Rosetta spacecraft detected plasma regions with near zero magnetic field and relatively small fluctuations \citep{goetz_first_2016}. These regions were observed within a few hundred kilometers from the comet and appeared to be highly sporadic and transient. Spacecraft dwell time at each event varied from seconds to more than 40 minutes, indicating the very dynamic and variable size of these structures \citep{goetz_structure_2016,timar_modelling_2017}. Around the diamagnetic regions the bulk electron density is closely related to the local neutral density \citep{eriksson_cold_2017,henri_diamagnetic_2017,hajra_dynamic_2018}. \citet{henri_diamagnetic_2017} showed that a relation can be established between the electron exobase and the observed diamagnetic boundary distances. On the other hand, suprathermal electrons show a peculiar signature in which, at each crossing into the field-free regions, flux of electrons with energies from tens of eV to several hundreds of eV decreases \citep{madanian_plasma_2016,nemeth_charged_2016}.

Rosetta observations near perihelion mostly represent a combination of shocked, highly perturbed, and heated solar wind plasma, plus electrons and ions of cometary origin. Given that comet 67P has no intrinsic magnetic field \citep{auster_nonmagnetic_2015}, and that solar wind ions have been obscured far upstream \citep{nilsson_evolution_2017}, solar wind electrons play a critical role in carrying the IMF through the coma \citep{plaschke_first_2018}. Furthermore, the spacecraft distance to the comet at this time is only a few hundred kilometers which is comparable to or smaller than ion gyroradii. Therefore, studying the small scale electron dynamics is crucial in understanding the nature of these events. Explicitly, electrons around the comet can originate from three sources: solar wind electrons, photoelectrons, and secondary electrons from electron-impact ionization \citep{galand_ionospheric_2016,madanian_suprathermal_2016,vigren_model-observation_2016, heritier_plasma_2018}. Models of electron production around comet 67P at perihelion have shown that without acceleration processes, photoionization is the main source of electron production up to about 70 eV, while at higher energies solar wind electrons become dominant \citep{madanian_plasma_2016}. Unperturbed solar wind suprathermal electrons exhibit distinct non-Maxwellian features; an isotropic component known as “halo”, and a field-aligned “strahl” beam propagating usually in the anti-sunward direction \citep{feldman_solar_1975}, though in the turbulent plasma environment of the inner coma and near diamagnetic cavities, these distributions will likely be modified.

In this paper we investigate the electron dynamics near diamagnetic regions of comet 67P. We show how bulk electron densities change across diamagnetic boundaries. We analyze the energy extent of suprathermal electron flux difference between inside and outside the cavities. We provide a detailed case study on how suprathermal electron pitch angle distributions (PADs) evolve, which has implications for the energy range of flux differences and the size of the cavities. In section 2, we describe the instruments and data processing method. Our observations are presented in section 3. We discuss the results and review the interpretations in section 4 and finally, provide our conclusions in section 5.

\section{Instrumentation and Data Processing Method}
We use data from the Magnetometer (MAG) \citep{glassmeier_rpc-mag_2007}, the Ion and Electron Sensor (IES) \citep{burch_rpc-ies:_2007}, the Langmuir Probe (LAP) \citep{eriksson_rpc-lap:_2007}, and Rosetta Orbiter Spectrometer for Ion and Neutral Analysis / Cometary Pressure Sensor (ROSINA/COPS) \citep{balsiger_rosina_2007} in our analysis. The first three instruments are part of the Rosetta Plasma Consortium \citep{carr_rpc:_2007}. In the following, we describe the IES instrument in detail and provide a brief description of the other instruments. For more details on each instrument, readers are referred to the corresponding instrument papers and references therein. Three electron sensors onboard the Rosetta spacecraft with different detection methods have enabled us to study the electron dynamics at different energy ranges through the perspective of each instrument. Measurements of LAP and Mutual Impedance Probe (MIP) \citep{trotignon_rpc-mip:_2007} instruments describe the bulk electron population, while IES measures electrons across a wide range of energies. There is no specific set of criteria to categorize the electron populations to our knowledge, and different authors have chosen different energy ranges to label cold, thermal, suprathermal, and in some cases warm electron populations. In this paper, the term suprathermal electron refers to IES electron measurements with energies above $\sim$10 eV, and electrons below this energy constitute the bulk population. We use LAP data to estimate bulk electron densities. For electron directional variability analysis, we are only interested in the suprathermal electrons that have energies exceeding 100 eV. 

\subsection{IES}
The IES instrument on Rosetta is capable of measuring near full 3-D distribution of charged particles \citep{burch_rpc-ies:_2007}. The IES consists of two stacked toroidal electrostatic analyzers that measure electrons and ions with energies between 4.3 eV and 18 keV. The IES energy resolution is 8 percent at each energy bin and the instrument has a $360^{\circ}$ azimuthal by $90^{\circ}$ polar field of view, providing $2.8\pi$ solid angle coverage. The electron sensor, which is emphasized in this paper, has a $22.5^{\circ}$ azimuthal resolution provided by 16 anodes. It was initially designed to scan the polar coordinate through 18 elevation steps with a $5^{\circ}$ resolution. Later on, due to engineering reasons the in-flight software was modified to scan 16 elevation steps with $6^{\circ}$ between each step. Note the different angle nomenclature in \citet{burch_rpc-ies:_2007}.

The IES instrument was mounted on the corner of the spacecraft providing a perfect pointing during most of the mission for probing the solar wind. For the period near perihelion, IES measurements in every two adjacent energy channel pairs, elevation step pairs, and azimuthal anode pairs were averaged onboard before transmission to fit the available telemetry rate. In our analysis, we used all individual sectors in the IES field of view (FOV) by assuming that the paired sectors share the averaged value evenly. We should note that soon after the beginning of the mission, two IES anodes (anodes 11 and 12) became malfunctional and did not return reliable data while their neighboring anodes performed nominally. Furthermore, in early April 2015, the instrument showed reduced count intensities in half of the anodes (anodes 8-15). These anodes shared the same octal amplifier. Further analysis of data showed that this decrease in amplification efficiency affects primarily low energy bins that also experienced saturation at high count rates. High energy bins ($\sim$100 eV and above) are not affected. We addressed these issues in our calibration analysis and considered the possible induced uncertainties before drawing conclusions.

To convert IES raw counts to a physical parameter such as differential electron flux, we rely on the instrument geometric factor per sector of $G = 3\times10^{-5}$ cm\textsuperscript{2} str eV/(eV counts/electron) \citep{burch_rpc-ies:_2007}. This geometric factor must be updated by appropriate correction factors to account for spacecraft blockage and change in the instrument’s microchannel plate detection efficiency. Measurements inside a 20 minute diamagnetic region on 26 July 2015 were used for calibration. Our assumption here is that the electron gas inside the diamagnetic cavity is isotropic. Counts in anodes 0--7 and elevation steps 5--15 were averaged over time to obtain a nominal isotropic count value per energy step. These sectors are free from spacecraft blockage, include the Sun viewing FOV, and anodes are considered healthy. Next, for every sector a correction factor was calculated by taking the ratio of the sector count to the isotropic count. A similar method of in-flight calibration was employed by \citet{broiles_characterizing_2016} in an earlier stage of the mission using solar wind data. The particle differential flux in anode $i$ and elevation $j$ measured at energy step $k$ is determined from \citep{madanian_suprathermal_2016,broiles_characterizing_2016}:

\begin{equation}
J_{ijk} = \frac{\dot{c}_{ijk}}{G*\chi_{ijk} *E_{k}} 
\end{equation}

\noindent where $\dot{c}_{ijk}$ is the count rate, $\chi_{ijk}$ is the new correction factor, and $E_{k}$ is the energy. For tables of $\chi_{ijk}$ see Section \ref{sec:data}. Employing the new correction factors also improved the low amplification rates at low energies. Figures \ref{figGcompare_IESFOV} and \ref{figGcompare_totflx} in the appendix section show a comparison of the fluxes based on different geometric factors. We consider IES background noise to be small and we did not subtract a constant background rate from data. Energy spectra are shifted in energy to correct for the spacecraft potential. For a negative (positive) potential, fluxes are shifted to higher (lower) energies. The spacecraft potential can also deflect electron trajectories; however, for typical spacecraft potentials near perihelion ($\sim -10$ V), deflection of high energy electrons ($E > \sim100$ eV) is very low and negligible \citep{scime_effects_1994}.

\subsection{MAG}
We used calibrated magnetic field data from the MAG instrument on the Rosetta spacecraft \citep{glassmeier_rpc-mag_2007}. The MAG instrument has two 3-axis fluxgate magnetometers mounted on a 1.5 m boom with 15 cm separation in between them.  Entering the magnetic field-free regions in April 2015 provided an opportunity to recalibrate the sensors and modify the temperature model for that period \citep{goetz_first_2016}. Our analysis is restricted to time periods for which calibrated magnetic field data are available. The magnetic field data are low band pass filtered to one second temporal resolution.

\subsection{LAP}
The LAP instrument consists of two spherical Langmuir probes, LAP1 and LAP2, mounted on two booms extending 2.2 and 1.6 m from the spacecraft, respectively \citep{eriksson_rpc-lap:_2007}. The instrument sweeps through voltage biases across the probes and the spacecraft ground to retrieve current-voltage curves that are subsequently used to derive plasma density and temperature, and spacecraft potential. LAP densities have been cross calibrated by corresponding MIP measurements and the data product has a variable time resolution less than three minutes.

\subsection{ROSINA/COPS}
The two pressure gauges on the ROSINA/COPS instrument provide density measurements of the neutral coma \citep{balsiger_rosina_2007}. The nude gauge density measurements have been adjusted for a water dominated coma which is expected near perihelion \citep{heritier_ion_2017,lauter_surface_2018, gasc_sensitivity_2017}. Neutral densities in the coma are mostly smooth and change on the timescale of the comet\textsc{\char13}s rotation.

\begin{figure}[H]
\centering
\includegraphics[width=15cm]{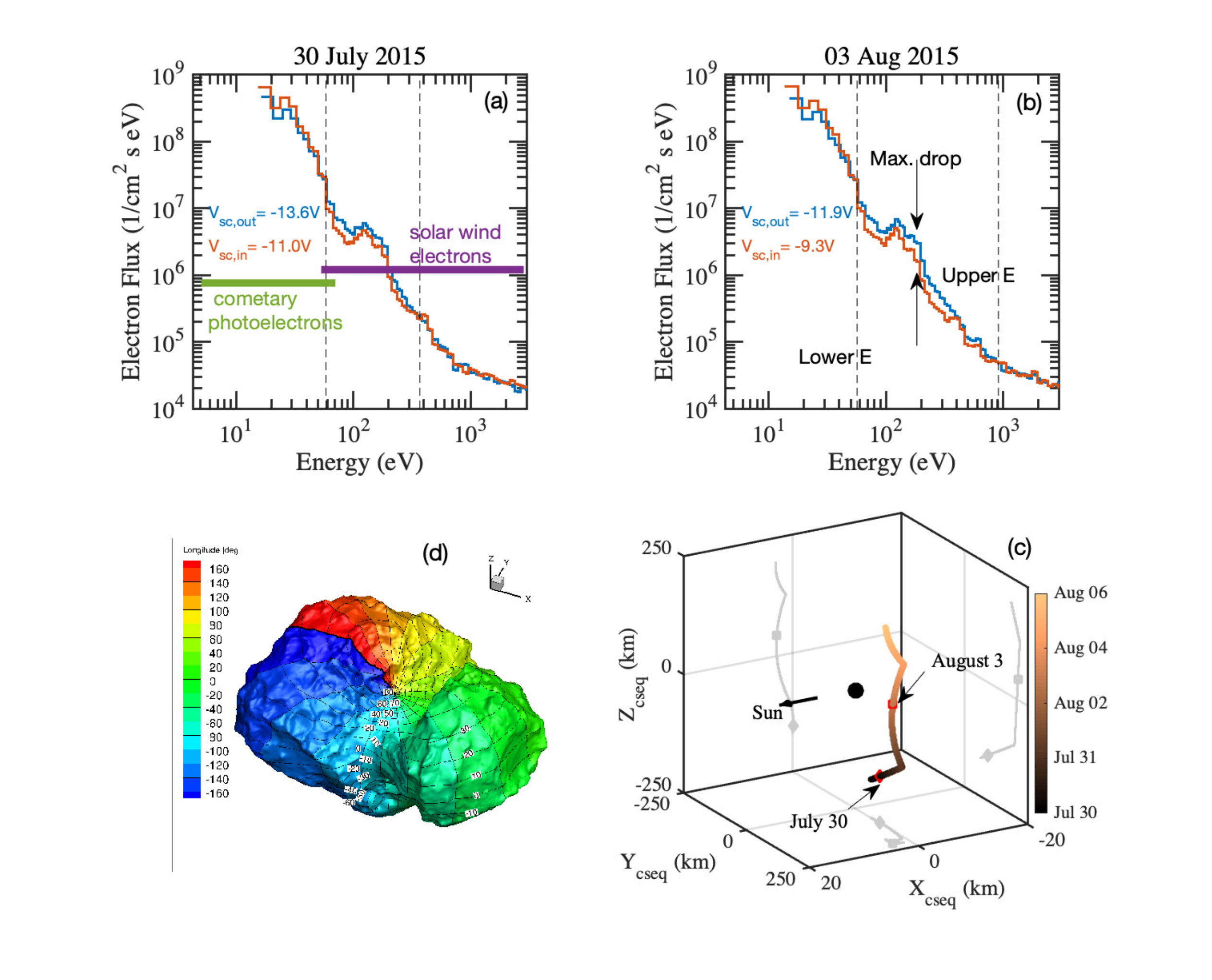}
\caption{Differential electron flux energy spectra integrated over the entire IES FOV inside (red) and outside (blue) the diamagnetic regions on (a) 30 July 2015 and (b) 3 August 2015. The corresponding spacecraft potentials are annotated on these panels. Error estimates due to the counting statistics are smaller than the line widths at most energies and are not shown. The horizontal green and purple lines on panel (a) highlight energy ranges of dominant cometary and solar wind electrons, respectively. (c) The Rosetta spacecraft trajectory around the comet. (d) Surface map of the comet illustrating latitudes and longitudes in the ESA/RMOC frame.}
\label{figexhibeves}
\end{figure}

\section{Observations}
The reduced electron flux inside diamagnetic regions has been discussed in a couple of studies \citep{madanian_plasma_2016,nemeth_charged_2016,timar_modelling_2017}. Figure \ref{figexhibeves} shows two examples of IES electron spectra inside and outside diamagnetic regions, each exhibiting a decreased flux over different energies. The top-left panel shows an event on 30 July 2015 and the top-right panel shows an event few days later on 3 August 2015. The ordinate axis in these plots represents the differential electron flux integrated over the entire IES FOV ($2.8\pi$ solid angle). The red (blue) lines on the top panels show the time averaged spectra inside (outside) the diamagnetic cavities. The period inside the diamagnetic cavity on 30 July is from 11:00:52 to 11:11:40 UTC and on 3 August from 17:20:42 to 17:28:03 UTC. The outside periods for 30 July and 3 August are selected between 10:52:26 -- 11:00:46 UTC and 17:12:16 -- 17:20:36 UTC, respectively. The horizontal green and purple lines on panel (a) are shown as references to highlight energy ranges of dominant cometary and solar wind electrons \citep{madanian_plasma_2016}. The vertical dashed lines show the energy range in which a flux difference is observed (lower and upper energies). 

On 30 July, flux of electrons in the $\sim60 - 350$ eV range inside the diamagnetic region has decreased by variable amounts. This energy range extends to around 900 eV on 3 August. A characteristic energy indicated by \textsc{\char13}Max. drop\textsc{\char13} at around 175 eV on panel (b) is the energy at which the highest flux difference is observed. This energy for the event on 30 July is around 74 eV. Panel (c) in Figure \ref{figexhibeves} shows the Rosetta spacecraft trajectory around the comet between 30 July and 6 August 2015. The colorbar represents the time. The reference frame in this plot is the dynamic body-Centered Solar Equatorial (CSEQ) frame in which the $+x$ axis is toward the Sun, the $+z$ axis is aligned with the projection of the solar rotation axis on a plane perpendicular to the $x$ axis, and the $y$ axis completes the right-hand coordinate system. The frame\textsc{\char13}s origin is the comet\textsc{\char13}s center of mass (shown with a black dot). Rosetta was at around 180 km from the comet on 30 July, and it gradually moved to a distance of 250 km north-east of the comet on 6 August. The spacecraft speed with respect to the comet was a few meters per second. We will discuss latitudinal dependence of variables. The latitude is measured in the ESA/RMOC shape frame (also known as the landmark coordinates) illustrated in the surface map of the comet in panel (d). Colors represent different longitudes, while latitudes are annotated on the map.

The flux difference across the diamagnetic boundaries creates an energy density difference between inside and outside plasmas. As seen in Figure \ref{figexhibeves}, the energy range of flux difference varies for different diamagnetic events, and this variability has not been studied so far. In Section 3.1 we provide a statistical analysis of a subset of diamagnetic events and in Section 3.2 a detailed case study for one of these events is presented.

\subsection{Analysis of Suprathermal Electron Flux Difference across Diamagnetic Boundaries}

We use a subset of diamagnetic events reported in \citet{goetz_first_2016} and limit our study to July and August of 2015, when comet activity was relatively high and the majority of diamagnetic events were observed. With the IES measurement cycle in mind, we down-selected events lasting longer than 256 s and with at least 512 s separation from another event on at least one side. These criteria ensure that at least one full IES measurement cycle exist inside the diamagnetic region and that the outside measurements are not contaminated by shorter events. This brought down the number of events from a total of 313 to 62 events. For the list of events see Section \ref{sec:data}. We used an algorithm to search and compare the IES energy spectra inside and outside each event and record energy bins with reduced electron fluxes. For 31 events we had the option to choose the outside spectrum from the trailing or the leading side. For these cases measurements from the side with the higher magnetic field strength were selected. For events that showed multiple drops corresponding to multiple energy ranges (i.e., the inside spectrum would drop below the outside spectrum multiple times due to similar overlapping spectra,) the widest energy range was recorded. We present our observations in the context of total energy flux difference, $\Delta\psi$, across boundaries as seen in IES electron spectra and defined by:
\begin{equation}
\Delta\psi = \sum_{k=E_{lower}}^{E_{upper}} (\psi(E_k)_{out} - \psi(E_k)_{in})\times E_k
\end{equation}

 \begin{figure}[H]
 \centering
 \includegraphics[width=0.9\linewidth]{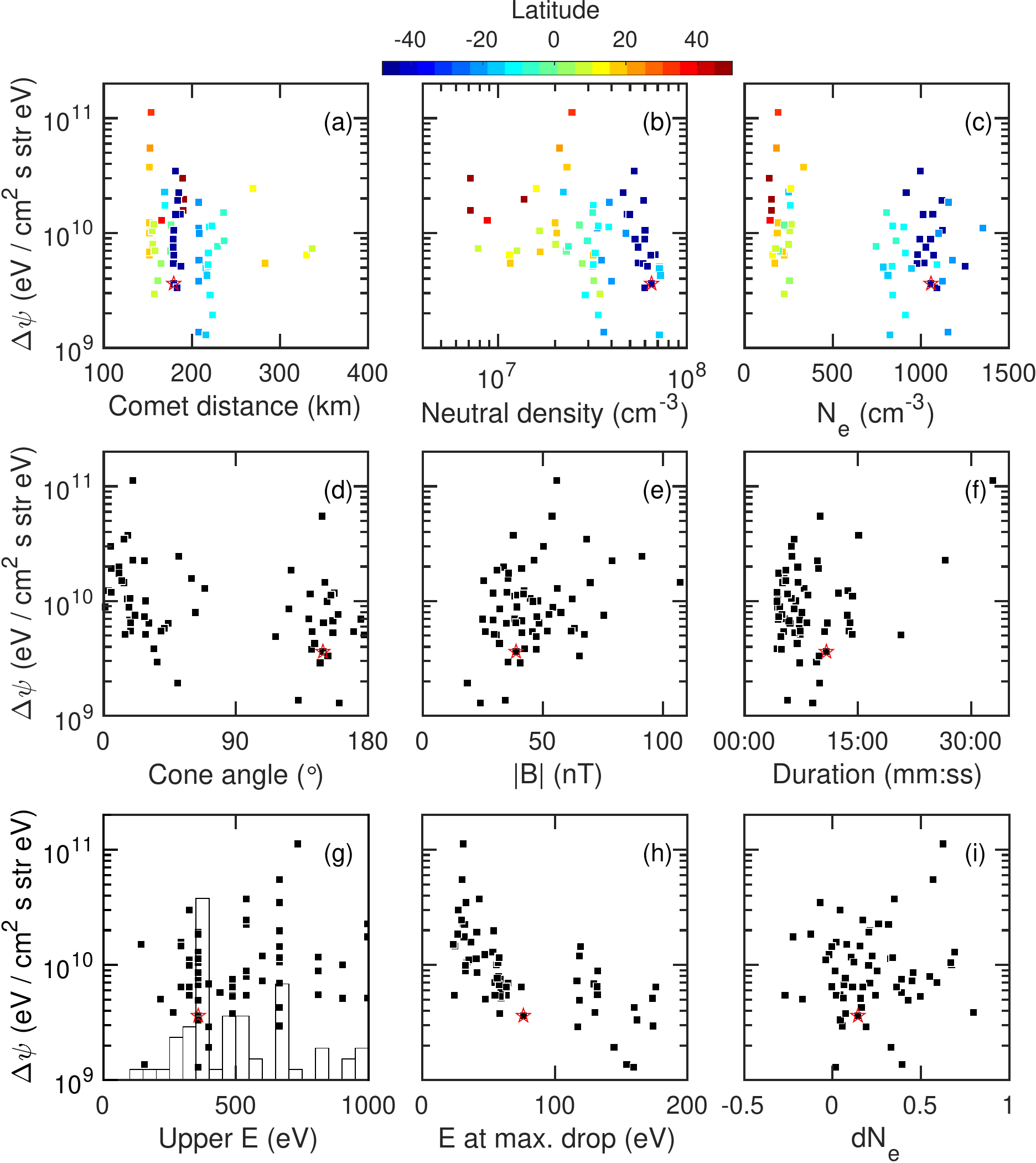}
 \caption{Distributions of plasma parameters for 62 diamagnetic events included in the study. The ordinate axis in all panels shows the parameter $\Delta\psi$. Panels (a - c) show the distributions of cometocentric distance, local neutral density, and bulk electron density, respectively. Data in these panels are also color coded by the cometary latitude. Panels (d - f) show distributions of cone angle, magnetic field strength, and event duration. Panels (g - i) show the distributions of upper energy limit of flux variations, energy of the highest flux difference, and bulk electron density difference. The red star marks the event on 30 July 2015 which is considered further in Section \ref{sec:30julyeve}.}
\label{figstatmultipanel}
\end{figure}

\noindent where $\psi(E_k)$ is the integrated differential electron flux over the IES FOV at energy $E_k$. $\Delta\psi$ distributions against several other parameters are presented in Figure \ref{figstatmultipanel}.  The first row in this figure shows $\Delta\psi$ as a function of cometocentric distance, neutral density and bulk electron density $N_e$ measured by LAP instrument, respectively. These panels are also color coded based on the cometary latitude at each event (see Figure \ref{figexhibeves}, panel (d)). As shown in panel (a), observations are mostly within 300 km from the comet and neutral densities varies between $5\times10^6 - 10^{8}$ cm\textsuperscript{-3}. The neutral densities also show a clear latitudinal dependence. Data in panel (b) shows that the comet is significantly more active in the southern hemisphere \citep{hansen_evolution_2016,hassig_time_2015,lauter_surface_2018}. During the perihelion passage, the southern hemisphere of the comet receives higher insolation and this period is in the midst of the southern summer in cometary seasons \citep{keller_insolation_2015}.

Panel (c) shows LAP electron densities measured at the beginning of each diamagnetic crossing. The LAP densities are clustered around 200 cm\textsuperscript{-3} and 1000 cm\textsuperscript{-3}. The higher density cluster, corresponding to events over the southern latitudes where the comet activity is higher, shows more variations. The lower density events around 200 cm\textsuperscript{-3} are more contained and show less variations. A few points that exhibit the highest $\Delta\psi$ are within this group. Since neutral densities show gradually increase with respect to decreasing latitude, one would expect to see a gradual increase in LAP electron densities at lower latitudes. However, there is a distinct separation in electron densities measured in the southern versus northern latitudes. This may reflect that the bulk radial plasma velocity is higher on the less active side. A reason for this could be that ion-neutral collisions, on the less active side, occur less frequently and thus are less efficient in hampering ion-acceleration along an ambipolar electric field (e.g., \citet{vigren_1d_2017}). In addition, in a simplified view, an equally pronounced outward radial acceleration on the more active side would conflict with momentum conservation. Most events in the southern hemisphere occurred between 26 July and 3 August, when most of the long-lasting diamagnetic events have been observed. 

Panels (d - f) show, respectively, distributions of the magnetic field cone angle, magnetic field strength, and event duration. The cone angle defines the angle between the magnetic field vector and the comet-Sun line. The distribution in panel (d) shows events grouped around $30^{\circ}$ and $150^{\circ}$ cone angles which is expected for the observations near perihelion as significant magnetic field draping exists and the spacecraft resides mostly in the terminator plane at this time. Correlation between electron number flux and magnetic field magnitude slightly increases at higher energies when all measurements at perihelion are included \citep{madanian_plasma_2016}, but the $\Delta\psi$ distribution in panel (e) exhibits no or a very weak correlation with the magnetic field strength. Event durations varied between 257 seconds and 32 minutes. The longest event that also shows the highest $\Delta\psi$ is on 7 July 2015 at 09:44:22 UTC. The outside spectrum is selected from the trailing side of that event the flux difference extends up to 733 eV. 

The third row in Figure \ref{figstatmultipanel} shows $\Delta\psi$, respectively, versus the upper energy limit of flux difference, energy of the highest flux difference, and relative difference in bulk electron density between inside and outside plasmas, $dN_e = (N_{e_{out}} - N_{e_{in}})/N_{e_{out}}$. The histogram in the background of panel (g) shows the occurrence rate of the upper energy limit. Although upper limits spread across many energies, the distribution suggests that the flux decrease stops at certain energies more often. The first peak in the histogram at 350--400 eV bin is the most dominant and includes 17 events. Flux difference for nine events extends up to 650--700 eV (the second highest peak). We will revisit this point in Section 3.2. $\Delta\psi$ decreases when the most affected electrons are at higher energies which can be observed in panel (h). In addition, panel(i) shows that for most events bulk electron density inside the diamagnetic region decreases, confirming previous findings using MIP data \citep{henri_diamagnetic_2017}, though this decrease shows no apparent relation with IES flux differences. Suprathermal electrons at 100 or 200 eV travel through the plasma at speeds significantly faster than bulk electrons. Their flux variability occurs on time scales much different than the bulk plasma variation observed inside diamagnetic regions \citep{hajra_dynamic_2018}.

\subsection{Suprathermal Electron PADs Case Study for the Event on 30 July 2015, 11:00:51 UTC} \label{sec:30julyeve}
The diamagnetic cavity event that we consider in this section was shown in Figure \ref{figexhibeves} panel (a). It is observed at negative latitudes and is one of the 17 events for which flux difference extends to $\sim$350 eV (see panel (g) in Figure \ref{figstatmultipanel}). Table \ref{table:1} lists plasma and field parameters around this event.

\begin{figure}[H]
 \centering
 \includegraphics[width=0.95\textwidth]{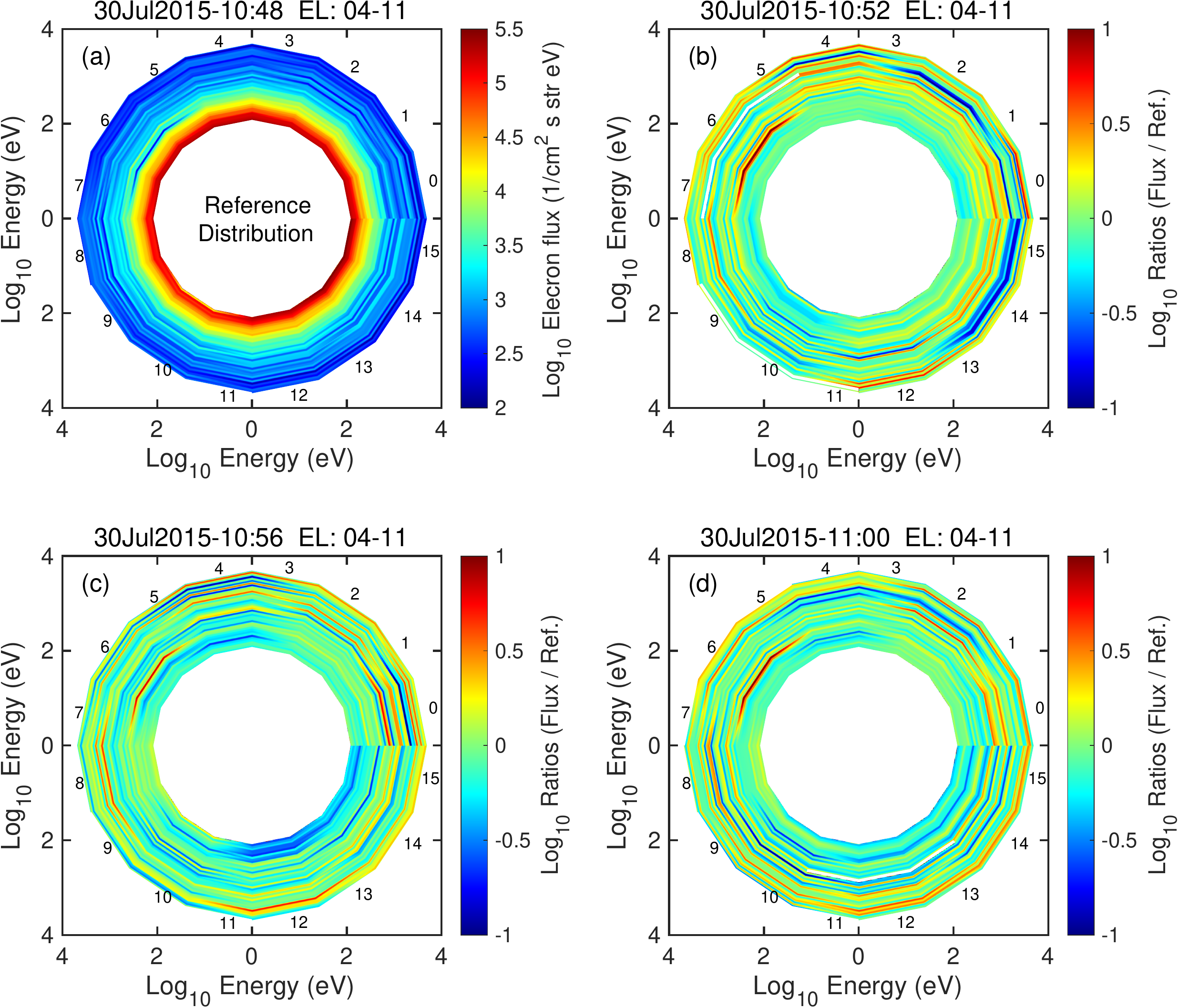}
 \caption{2D cuts of the IES FOV showing electron differential flux variations in four timestamps between 10:46 and 11:00 UTC before the diamagnetic cavity crossing on 30 July 2015. Panel (a) shows the electron differential flux at the first timestamp. Panels (b - d) show the corresponding flux ratios with respect to the distribution in panel (a).}
 \label{figpolarplotsJim}
  \end{figure}

\begin{table}[!h]
\centering
\caption{Plasma and field parameters for diamagnetic cavity event on 30 July 2015} 
\begin{tabular}{p{6cm}p{2cm}p{2cm}}
\hline 
$r_{comet}$ (km) & \multicolumn{2}{l}{179.5} \\ 
$D_{sun}$ (AU) & \multicolumn{2}{l}{1.24} \\ 
Neutral density (cm\textsuperscript{-3}) & \multicolumn{2}{l}{$6.7\times10^{7}$} \\ 
Latitude  & \multicolumn{2}{l}{-48} \\
Cone angle  & \multicolumn{2}{l}{149.3} \\
B (nT) & \multicolumn{2}{l}{38.8} \\
Duration & \multicolumn{2}{l}{00:10:55 (11:00:51 - 11:11:41 UTC)} \\
Energy range of reduced flux (eV) & \multicolumn{2}{l}{56.1 - 358} \\
Energy of max. flux difference (eV) & \multicolumn{2}{l}{74.4} \\
  & Inside & Outside \\
 LAP bulk electrons density (cm\textsuperscript{-3}) & 997.3 & 1164.8 \\
\hline 
\end{tabular}
\label{table:1}
\end{table}
To better understand the nature of the reduced fluxes during the transition into the diamagnetic region we examine the 3D spatial distributions of high energy suprathermal electrons. Figure \ref{figpolarplotsJim} shows 2D cuts of electron distribution variations in the IES FOV for four timestamps before the diamagnetic event on 30 July 2015. Panel (a) shows the differential electron flux for IES anodes (labeled 0-15) averaged around the central elevation plane at the first timestamp and is labeled as the "reference" distribution. The colors are in logarithmic scale and energies between 100 eV and 5 keV are shown. Panels (b - d) show the flux ratios in the next three timestamps (all still outside the cavity) as compared to the reference distribution. The disconnection at 3 o\textsc{\char13}clock on these panels is an artefact of the plotting software.

Relative enhancements (red segments) are observed in anodes 0, 6, 8, and 12 of panels (b), (c), and (d); while decreases (blue segments) occur in anodes 2, 14, and 15 of panels (b) and (d) and in anodes 4, 6, 12 of panel (c). From this figure we notice directional changes for electrons at different energies close to the diamagnetic cavity. It is important to consider these changes in the electron trajectory with respect to the magnetic field. To better analyze these spatial changes, we analyze the electron pitch angle distributions. 

We should note that electron PAD is not an official data product of the IES instrument. Few factors that may complicate derivation of PADs and limit our ability to interpret them include, (1) low time resolution in IES data does not allow to resolve plasma effects such as wave-particle interactions in the distributions, (2) IES FOV does not cover the full sky and if the magnetic field points toward these gaps in the FOV (i.e. instrument symmetry axis,) part of the distribution will be lost, and (3) IES onboard averaging can reduce the resolution of the derived PADs. It is not our intention to study fine timescale effects on electrons, but rather we are looking at effects of changing magnetic field topology and our results prove that PADs at the current resolution can provide valuable information about those effects. We inspected the IES FOV for pitch angle coverage and ensured that the magnetic field direction during this event is favorable for PAD analysis.

The IES time resolution for a full cycle in the current mode is 256 s, resulting in a 2 s sampling time per energy bin. At each energy step, the deflector plates are biased in a see-saw fashion to conserve power and reduce sweep time. We track the time at which different energies and sectors were scanned within a cycle and update the magnetic field vector accordingly before calculating the pitch angles. An array consisting of 12 bins, each $15^{\circ}$ wide, is used to sort fluxes into the pitch angle space. To account for straddling of sectors that covered more than one pitch angle bin, sector flux is distributed across all overlapping bins and the final PADs are normalized by the sampling rate at each bin. 

\begin{figure}[H]
 \centering
 \includegraphics[width=0.9\linewidth]{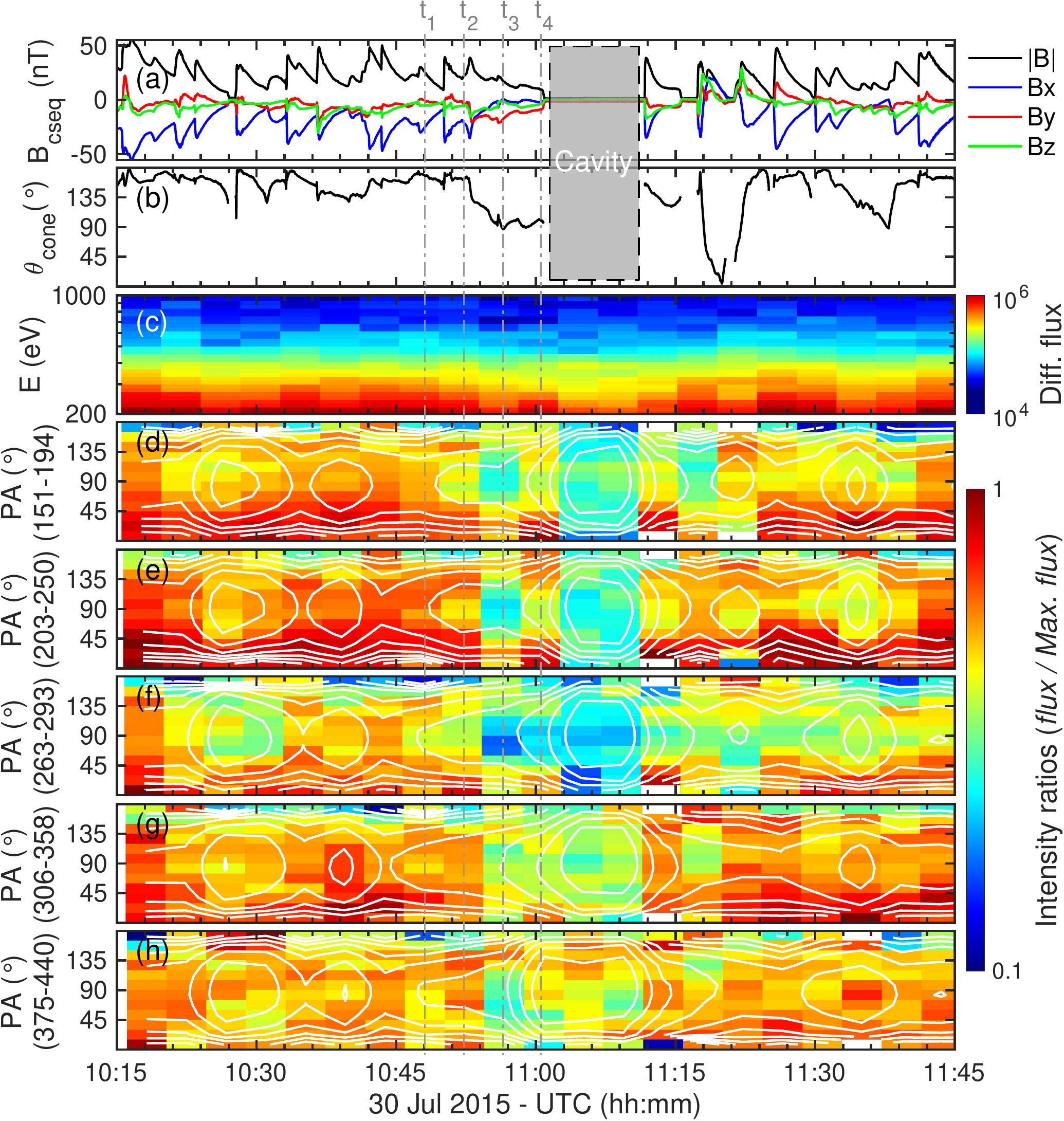}
 \caption{Magnetic field and electron distribution time series around the diamagnetic cavity on 30 July 2015. The field free cavity is observed between 11:00:51 and 11:11:41 UTC and is marked with a grey box. Panel (a) shows magnetic field components and magnitude in CSEQ coordinates, panel (b) shows the magnetic field cone angle, panel (c) is the differential electron flux spectrogram in units of $\log_{10}$(cm\textsuperscript{2} s eV)\textsuperscript{-1}, and panels (d - h) show electron pitch angle distributions in five different energy ranges. The fluxes are normalized by the maximum flux value in each panel. The white lines on these panels are the contours of the constant adiabatic invariant. The vertical dashed-dotted lines mark four IES timestamps before the onset of the diamagnetic cavity.}
\label{figPADtimeseries}
\end{figure}

The event on 30 July 2015 at 11:00:51 UTC meets our selection criteria. Specifically, we searched for periods of gradual changes in magnetic field strength over a few consecutive IES timestamps, where high amplitude magnetic field fluctuations were relatively low, as they can modulate the distribution faster than the IES can record and therefore cannot be studied. For the event studied in this section, although we do not observe the typical signatures of ultra-low frequency (ULF) waves, or circularly polarized whistler waves (see panel (a) of Figure \ref{figPADtimeseries}), we have to assume that wave-particle interactions are negligible.

Figure \ref{figPADtimeseries} shows an overview of magnetic field data and electron PADs across this event. The top panel in this figure shows the magnetic field components and magnitude in the CSEQ coordinates. The diamagnetic cavity event is identified between 11:00:51 and 11:11:41 UTC. The cone angle ($\theta_{cone}$) is shown in panel (b). The spectrogram in panel (c) shows the FOV integrated differential electron flux (cm\textsuperscript{2} s eV)\textsuperscript{-1} as a function of energy in the 200-1000 eV range. Flux reductions inside the cavity for this event were previously illustrated in panel (a) of Figure \ref{figexhibeves}, and can also be identified in panel (c). Panels (d - h) show the electron PAD time series at different energies normalized by the maximum flux value in each panel. The distributions have been averaged over consecutive energy bins to improve the counting statistics. The energy ranges are specified in the parentheses. All colorbars are in logarithmic scale. The white lines overplotted on these panels are contours of constant magnetic moment, $\mu_m = W_{\perp} / |B|$, where $|B|$ is the magnetic field magnitude and $W_{\perp} = 1/2 \mbox{ } m_e V_{\perp}^2$ is the perpendicular energy of electrons. The pitch angle distributions and contours inside the cavity have no physical meaning.

Between 10:53:00 and 11:00:00 UTC, the magnetic field shows, on average, a gradual decrease in the field strength. There are perturbations due to the turbulent plasma environment. The $B_x$ component is shown with the blue color in panel (a) of Figure \ref{figPADtimeseries}, and is highly negative throughout this period. In fact, most of the variations in the magnetic field strength originates from the $B_x$ component while the two other components are relatively quiet.  Close to the diamagnetic region the $y$ component of the field becomes dominant and shows a continuous decline. The magnetic field direction changes from anti-sunward (cone angle $\sim180^{\circ}$) to a direction perpendicular to the comet-Sun line (cone angle $\sim90^{\circ}$). This period corresponds to four IES timestamps identified by vertical dashed-dotted lines drawn across all panels and labeled by $t_1$, $t_2$, $t_3$, and $t_4$.

At 10:45 UTC electrons show a fairly scattered distribution occupying most of the pitch angle bins with similar intensities, except for the distributions in panels (f) and (g). In the next four timestamps, flux reductions around $90^{\circ}$ pitch angles are observed and accompanied by increased fluxes in directions parallel ($0^{\circ}$) and anti-parallel ($180^{\circ}$) to the magnetic field. This is indicative of a changing distribution from isotropic to field-aligned. The effect is particularly evident for $151 - 293$ eV electrons, while $306-358$ eV electrons exhibit this change in the last two timestamps before the cavity. The redistributed electrons seem to follow along the white contours of the first adiabatic invariant. In contrast, the distributions of $375-440$ eV electrons in panel (h) show a different pattern. With the exception of timestamp ($t_3$) where an enhanced anti-parallel flux is observed, distributions are relatively disordered and chaotic with respect to the onset of the diamagnetic cavity, or the adiabatic invariant contours. This  implies that the first adiabatic invariant is only conserved up to a certain energy. 

\begin{figure}[H]
\centering
\includegraphics[width=0.9\linewidth]{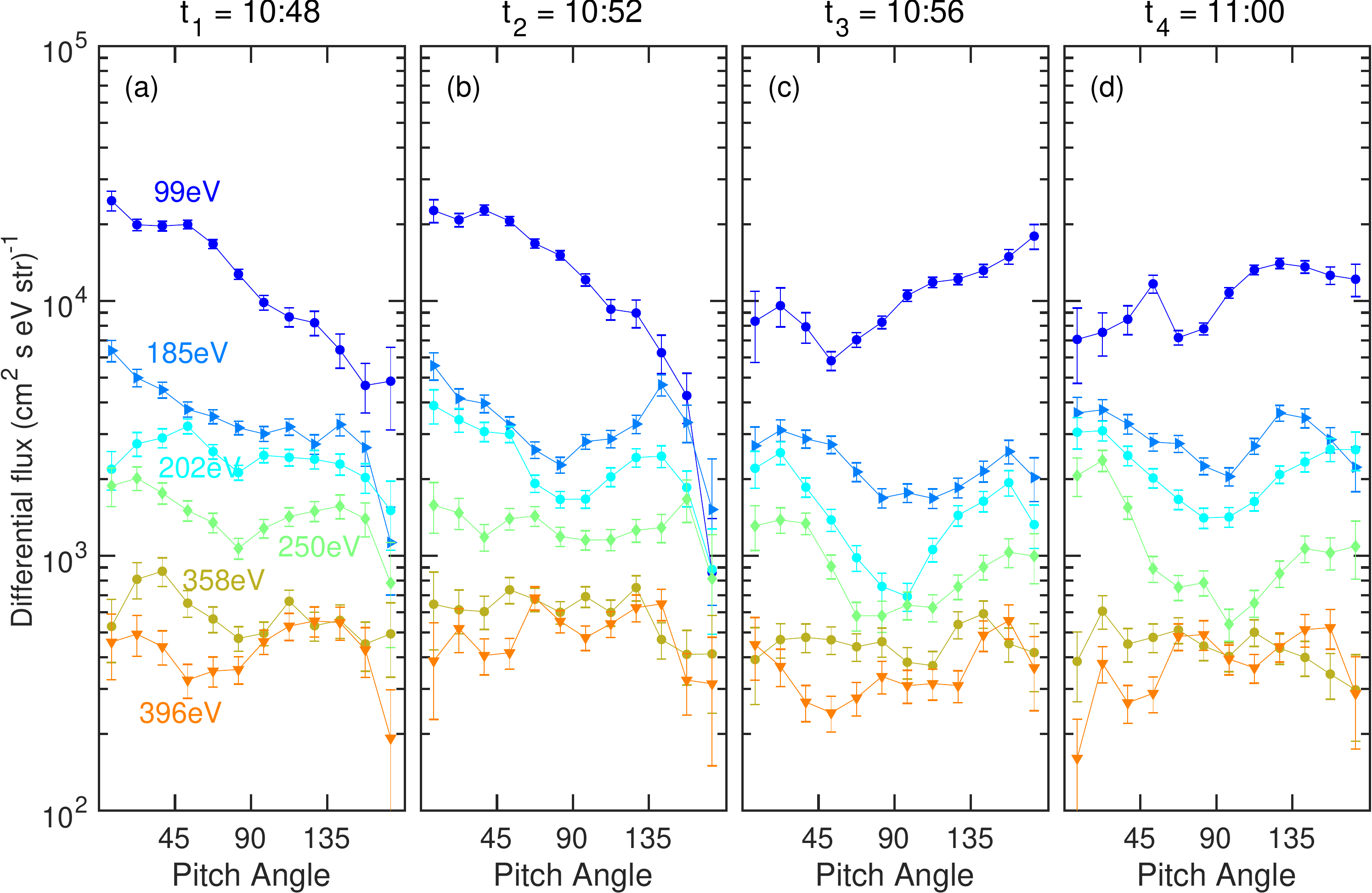}
\caption{Differential electron flux at selected energies (different colors) versus pitch angle at four timestamps prior to the cavity encounter on 30 July 2015. Energies are annotated in panel (a). Each panel correspond to a timestamp identified by vertical dashed-dotted lines in Figure \ref{figPADtimeseries}.}
\label{figlineplots}
\end{figure}

In Figure \ref{figlineplots} we examine these spectra in a more quantitative way. In panels (a - d), differential electron flux at selected energies are plotted versus pitch angle. Each panel corresponds to an IES timestamp marked with vertical, dashed lines in Figure \ref{figPADtimeseries}. The corresponding energies are annotated in panel (a), and error bars reflect the uncertainty due to the counting statistics. Error estimates for most data points are reasonably low and for a few points are larger. Larger error bars do not necessarily indicate that the observation must be discarded, but rather more measurements are needed to improve the confidence on the observation.

The 99 eV electrons have a maximum in the parallel direction until 10:52 UTC. In the next timestamp, the anti-parallel flux increases while the fluxes in $\sim0-80^{\circ}$ pitch angles decrease. Given that between 10:52 and 10:56 UTC the magnetic field vector rotates to mostly $-y$ direction, changes in 99 eV PADs indicate that a large flux of these electrons travel anti-parallel to the magnetic field and away from the comet. The 99 eV line also shows a sharp peak at around $50^{\circ}$ at 11:00 UTC. The 185 eV electron distribution in panel (a) shows a rapid fall in the last pitch angle bin. This is most likely due to the low count rate in that bin, as is evident by the larger error bars. The 185 eV distribution changes into a bidirectional, field-aligned pattern in the next three timestamps. Similarly, the 202 eV (cyan) and 250 eV (green) electrons start roughly isotropic and evolve into double-peak bidirectional distributions, while $90^{\circ}$ electrons become depleted. The net flux in these distributions remains almost the same from one timestamp to the next. In other words, the enhancements in pitch angles near $0^{\circ}$ and $180^{\circ}$ (as seen in panels (c) and (d)) are compensated by the depletions in $\sim90^{\circ}$ bins. The 358 and 396 eV lines, although moderately changing in time, do not show any depletion in perpendicular flux. The energies we discussed here are sensitive traces of the magnetic field topology. The depletion of $\sim90^{\circ}$ pitch angle electrons is consistent with adiabatic transport of electrons in sharply decreasing magnetic fields.

Spectra in timestamp $t_3$ show noticeably more depletion around $90^{\circ}$ pitch angles than the neighboring timestamps, but the net fluxes are still higher than those inside the cavity. The origin of this behavior has not been clearly identified at this time.

\section{Electron Dynamics, Discussion and Interpretations}
\subsection{Flux Difference and Implications for the Size of the Structures}
The magnetic field in the induced cometary magnetosphere is significantly higher than the IMF (50 nT in Figure \ref{figPADtimeseries} compared to a typical IMF $\sim$5 nT at 1 AU). The region of enhanced magnetic field creates a mirror point for solar wind electrons. Electrons that enter the cometary magnetosphere, mainly strahl electrons, will experience adiabatic heating and widening in the pitch angle distribution \citep{larson_electron_1992,kajdic_ninety_2014}. Our interpretation of the suprathermal electrons being of solar wind origin, particularly at energies considered in our pitch angle analysis, is based on model predictions, in which solar wind suprathermal electrons are the dominant population above $\sim$70 eV \citep{madanian_plasma_2016}. Though it should be noted that effects of acceleration processes were not included in the models. Also the differential fluxes shown in Figure \ref{figlineplots} are higher than the typical flux of solar wind halo electrons at 1 AU, but lower than the typical strahl component \citep{anderson_variability_2012,graham_evolution_2017}.

The suprathermal electrons that interact with the weakened magnetic field near diamagnetic regions are subject to adiabatic cooling and are redistributed to conserve the magnetic moment (see Figure \ref{figPADtimeseries}), resulting in decrease of the perpendicular flux and enhancement of fluxes along the magnetic field. The field-aligned electrons tied to the magnetic field will have limited access to the diamagnetic regions. For instance, prior to the diamagnetic event on 30 July 2015 (the four timestamps marked in Figure 4), electrons up to around 350 eV are effectively rearranged to field-aligned distributions. The flux of electrons in these energies is also reduced inside the cavity. Higher energy electrons ($E>350$ eV) do not exhibit field-aligned distributions in response to the decreasing magnetic field, and their intensity also remains unchanged during the passage of the diamagnetic cavity.

A possible explanation of this behavior could be associated with violation of the first adiabatic invariant beyond certain energies. It has been shown that in a region of highly curved magnetic field lines where the curvature radius is comparable to the electron gyroradius, resonance between the two leads to a significant scattering in electron gyration and aberration of the first adiabatic invariant \citep{buchner_regular_1989,young_magnetic_2008,zhang_first_2016}. In simple terms, the adiabatic invariant parameter $\kappa=\sqrt{R_c/R_L}$, where $R_c$ is the curvature radius and $R_L$ is the electron gyroradius, governs the energy threshold of the scattered electrons. Theoretical studies have shown that for $\kappa < 5$, curvature scattering occurs in almost all pitch angles and any anisotropic electron distribution becomes isotropic and chaotic \citep{buchner_regular_1989}. This prediction has also been confirmed by observations near ion diffusion zones in the Earth\textsc{\char13}s magnetotail where electrons interact with transient diamagnetic plasmas \citep{young_magnetic_2008,wang_multispacecraft_2010, wang_electron_2019}. 

It is natural to estimate the radius of the field line curvatures near the diamagnetic region based on this simple approach and the plasma parameters in Table 1. The gyroradius of a 350 eV electron in a 38 nT magnetic field is around 1.65 km, which implies a minimum curvature radius of $R_c\sim 40$ km. We speculate that the diamagnetic structure is at least 80 km wide and the shape is not necessarily symmetric. 

\subsection{Shape of the Interaction Region}
\begin{figure}[H]
\centering
\includegraphics[width=0.75\linewidth]{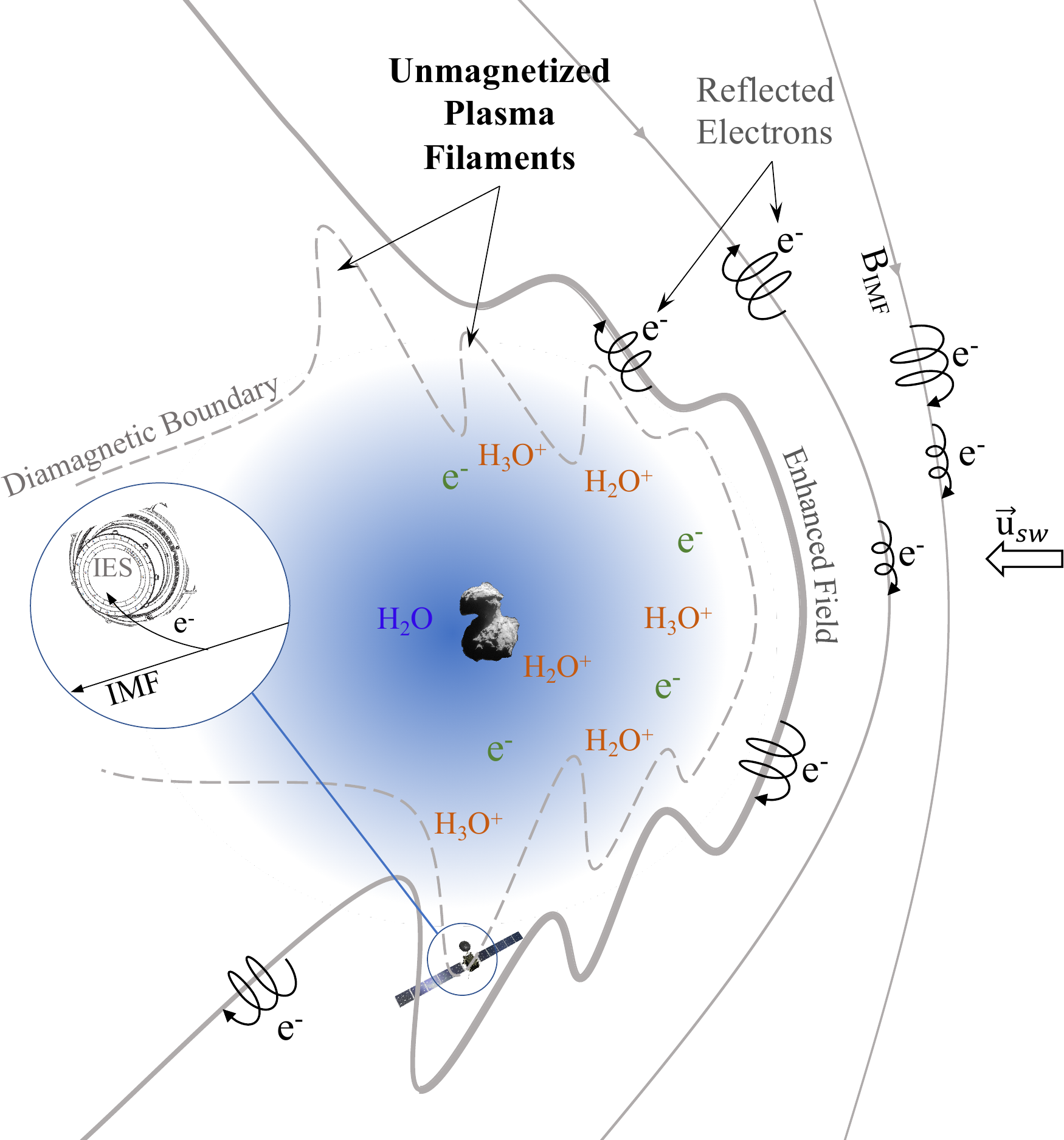}
\caption{Schematic illustration of a possible interpretation of Rosetta observations of diamagnetic regions around comet 67P, showing the electron dynamics and the asymmetrical expansion of unmagnetized plasma structures.}
\label{figschematic}
\end{figure}

Our estimate of the field line curvature around the diamagnetic regions is much smaller than the spacecraft distance to the comet, which suggests that the observed structures are in fact smaller than a global diamagnetic cavity surrounding the comet. Given the spacecraft\textsc{\char13}s relatively stationary position around the comet, and recurring diamagnetic encounters with different durations and sizes, it appears that what Rosetta observed were detached unmagnetized plasma clouds or filaments that convected over the spacecraft. Figure \ref{figschematic} shows a schematic illustration of a proposed scenario for the interaction region around the comet (also see Figure 5 in \citet{goetz_first_2016} and Figure 6 in \citet{henri_diamagnetic_2017}). The solar wind is incident from the right, the neutral coma is shown with a fading blue color, and the extent of the field free region is marked with the dashed line. The comet is surrounded by a region of enhanced magnetic field. The draped IMF lines are shown with solid, grey lines. The size of the cycloids next to electrons in this figure is a representation of the pitch angle. Solar wind electrons with larger pitch angles are reflected while small angle electrons reach the inner coma. The schematic in Figure \ref{figschematic} is a conceptual view to illustrate our findings and does not capture all aspects of the interaction such as microstructures within the boundary layer.

Fluid and hybrid models suggested that a global diamagnetic boundary can form at around 25 km, much closer to the comet that has been observed \citep{koenders_dynamical_2015,rubin_plasma_2014}. Simple one-dimensional MHD models \citep{cravens_physics_1986,ip_formation_1987}, although can predict the boundary distance at times, are not always consistent with the observations and models are in fact over simplified. For instance, it is assumed that ions are in chemical equilibrium, which is not always the case close to the boundary. Therefore, other mechanisms are perhaps in play that give rise to diamagnetic plasmas at the point of the spacecraft. Several authors have proposed different formation mechanisms \citep{henri_diamagnetic_2017,huang_possible_2016}. Kelvin-Helmholtz instabilities and transient snowplow structures have been proposed as processes that can accelerate unmagnetized ions parallel to the field downstream \citep{goetz_structure_2016,haerendel_dynamics_1986,koenders_dynamical_2015}. In this picture, the nonuniform neutral production and cometary ion outflow create an inhomogeneous plasma environment favorable for snowplow structures. Cometary ions are picked up and accelerated by the solar wind at different rates. Depending on the mass loading rate and solar wind motional electric field strength and orientation, ions can be accelerated in transverse directions, causing elongation of the structure boundary and the underlying outflowing unmagnetized plasma in a particular direction (see the unmagnetized plasma filaments in Figure \ref{figschematic}). These structures can also become detached from the core unmagnetized plasma \citep{halekas_plasma_2016}. Studying the evolution of these structures as they form upstream and propagate to the point of the spacecraft could provide more credence to this hypothesis and is left for a future study.

\subsection{Sources of Uncertainty}
We focused on events around perihelion when the comet neutral outgassing rate is significantly high. Yet, it is unlikely that collisional processes, mainly the electron-neutral collisions, undermine our results. Assuming an outgassing rate of $Q=10^{28}$ s\textsuperscript{-1} near perihelion, the total neutral density at $r = 200$ km from the comet is $n(r) = Q/(4\pi r^2 u_n) \approx 10^8$ cm\textsuperscript{-3}, where we assumed a spherically symmetric coma \citep{haser_distribution_1957} and a neutral outflow speed of $u_n = 1$ km/s. At 350 eV, the total (elastic and inelastic) electron-neutral collision cross section for water molecules is $\sim 3.5\times10^{-16}$ cm\textsuperscript{2}, resulting in a collisional mean free path, $\lambda_{mfp}$, of less than 300 km. For a 50 eV electron, $\lambda_{mfp}$ reduces to 100 km. Since these length scales are much larger compared to electron gyroradii, scattering due to electron-neutral collisions is inefficient at the energies considered. \citet{madanian_suprathermal_2016} discussed the effects of an ambipolar electric field that bounds suprathermal electrons to the inner coma, causing them to become further thermalized by collisional processes. The ambipolar electric field is fundamentally parallel to the magnetic field. However, if an ambipolar electric field was in play to accelerate electrons to energies as high as 250 eV in the parallel direction, we should should have seen a noticeable shift in energy spectra at lower energy electrons as well. Furthermore, the ambipolar electric field is a transient effect, and once charge separation occurs, electrons (ions) are decelerated (accelerated) to nullify the ambipolar field and retain charge quasi-neutrality. The fact that we see changes in PADs over several IES timestamps indicates that it is unlikely for high energy electrons to be driven by an ambipolar electric field.  

Another observation on the 30 July event is the presence of a peak in the 185 eV line at around $140^{\circ}$ in panel (b) or a peak in the 99 eV line at around $50^{\circ}$ in panel (d) of Figure \ref{figlineplots}, which can be interpreted as conic structures. The 202 and 250 eV lines in panels (c) and (d) of that figure, resemble butterfly distributions. Although several mechanisms (such as interaction with magnetosonic waves) can produce butterfly distributions, we should note that relatively larger error bars, for instance on the first three points of 99 eV line in panel (d), indicate that the distribution may be under-sampled on one side of the peak. Low angular resolution of the available data limits our ability to interpret finer details and PADs with large error bars must be viewed with caution.

\subsection{Role of Positive Ions}
The ion sensor on the IES measured noticeable amount of low energy pick up ions ($10<E<50$ eV) with very low but above background counts of higher energy ions up to around 3 keV. Near perihelion, Rosetta was inside a solar wind ion cavity \citep{nilsson_evolution_2017} and we did not observe any particular consistent pattern in ion energy distributions during the diamagnetic boundary crossings. More comprehensive analyses must be performed in the future using both the IES and the other electrostatic analyzer, the Ion Composition Analyzer (ICA) \citep{nilsson_rpc-ica:_2007}.

\section{Conclusions}
In this paper, we analyzed the dynamics of bulk and suprathermal electrons around comet 67P for a subset of long-lasting diamagnetic events. For events observed over the southern hemisphere bulk electron densities are noticeably higher than those over the northern side, even after taking neutral density and distance variations into account. Most events show lower bulk electron densities inside the diamagnetic region compared to the outside plasma \citep{henri_diamagnetic_2017}. Suprathermal electron fluxes inside the cavities show reductions over different energy ranges, most noticeably extending up to $\sim350$ eV for 27 percent of events and up to $\sim700$ eV for 14.5 percent of events. We present a first glance at suprathermal electron PADs at close proximity of an active comet. We propose a mechanism for interpreting PAD variations associated with the changing magnetic field topology, which is as follows: near diamagnetic regions, suprathermal electrons are transported adiabatically and rearrange from isotropic to field-aligned directions to conserve the magnetic moment. The field-aligned electrons tied to the magnetic field have limited access to diamagnetic regions, causing the observed decrease in electron flux between inside and outside the cavities. Electrons beyond certain energies behave non-adiabatically when their gyroradius is comparable to the magnetic field line curvature. These electrons do not show a systematic flux difference. For the diamagnetic event on 30 July 2015 at 11:00:51 UTC the size of the diamagnetic region is estimated to be at least 80 km. Why flux reductions for many events extend up to $\sim350-400$ eV is not completely understood yet and should be investigated further in future studies. 

We considered fairly long and isolated events when Rosetta spent few minutes inside each diamagnetic region, but there are shorter diamagnetic encounters that we did not discuss. We showed PADs in a generally decaying magnetic field environment. Small scale variations of the magnetic field can have an influence on PADs and further measurements and analyses are needed to verify the results shown in this paper.

Understanding the detailed behavior of electrons around comet 67P demands fully kinetic models with sufficient knowledge of the electromagnetic fields. The magnetic field structure around this comet is not a simple one. During some diamagnetic events the magnetic field showed a gradual decrease in strength before entering the diamagnetic plasma, but a sudden increase afterward. This asymmetry in the front and back envelopes requires further considerations. Analysis of suprathermal electron pitch angle distributions near perihelion and at other phases of the mission can provide another tool for studying plasma phenomena at this comet. Through Rosetta observations we now know that a cometary plasma environment is an exciting space plasma laboratory that deserves another visit not only to make improved plasma measurements, but to use new measurement techniques to study the evolution of plasma boundaries around an object with an inhomogeneous atmosphere.

\section{Acknowledgments}
Rosetta is a European Space Agency (ESA) mission with contributions from its member states and National Aeronautics and Space Administration (NASA). Work at Imperial College London was supported by STFC of UK under grant ST/N000692/1. EV is grateful for support from SNSA (Dnr 166/14). The work of ZN was supported by the ÚNKP-18-4 New National Excellence Program of the Ministry of Human Capacities and the Bolyai Janos Research Scholarship of the HAS. MR acknowledges the support of the State of Bern and the Swiss National Science Foundation (200020-182418).

\section{Data Availability} \label{sec:data}
All data used in this study can be accessed via the ESA Planetary Science Archive (\url{https://archives.esac.esa.int/psa}) as well as the NASA Planetary Data System (\url{https://pds.nasa.gov}). Tables of correction factors to the IES geometric factor and the list of events used in our statistical study are included as Supplementary materials.

\bibliography{Zoteroreferences.bib}

\appendix
\section{Comparing IES Fluxes with Different Geometrical Factors}
\renewcommand\thefigure{A.\arabic{figure}}    
\setcounter{figure}{0}    
\begin{figure}[H]
\centering
\includegraphics[width=0.7\linewidth]{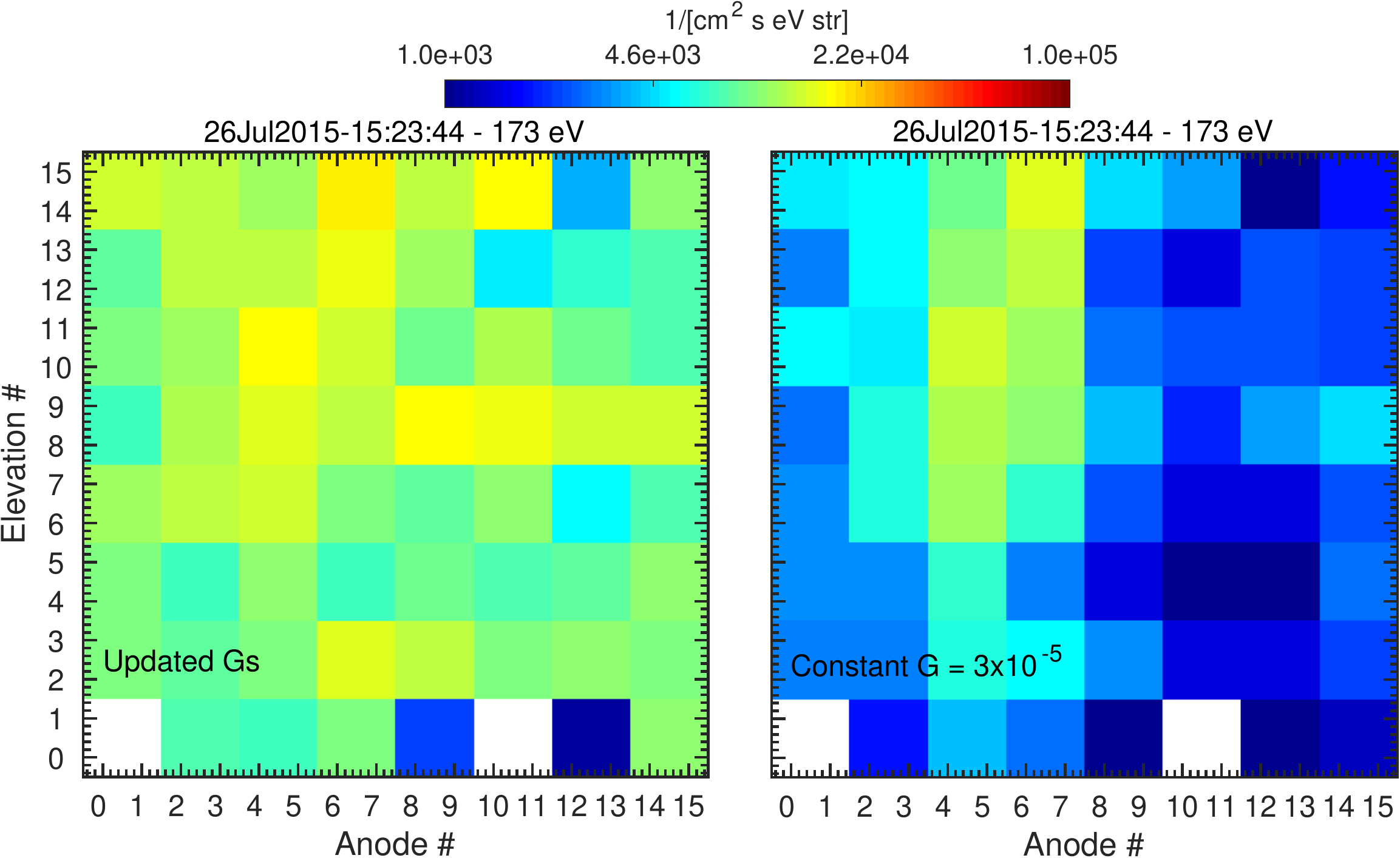}
\caption{Comparison of differential flux of 173 eV electrons in the IES FOV using the updated geometric factors (left) and the initial geometric factor of the instrument paper (right).}
\label{figGcompare_IESFOV}
\end{figure}

\begin{figure}[H]
\centering
\includegraphics[width=0.5\linewidth]{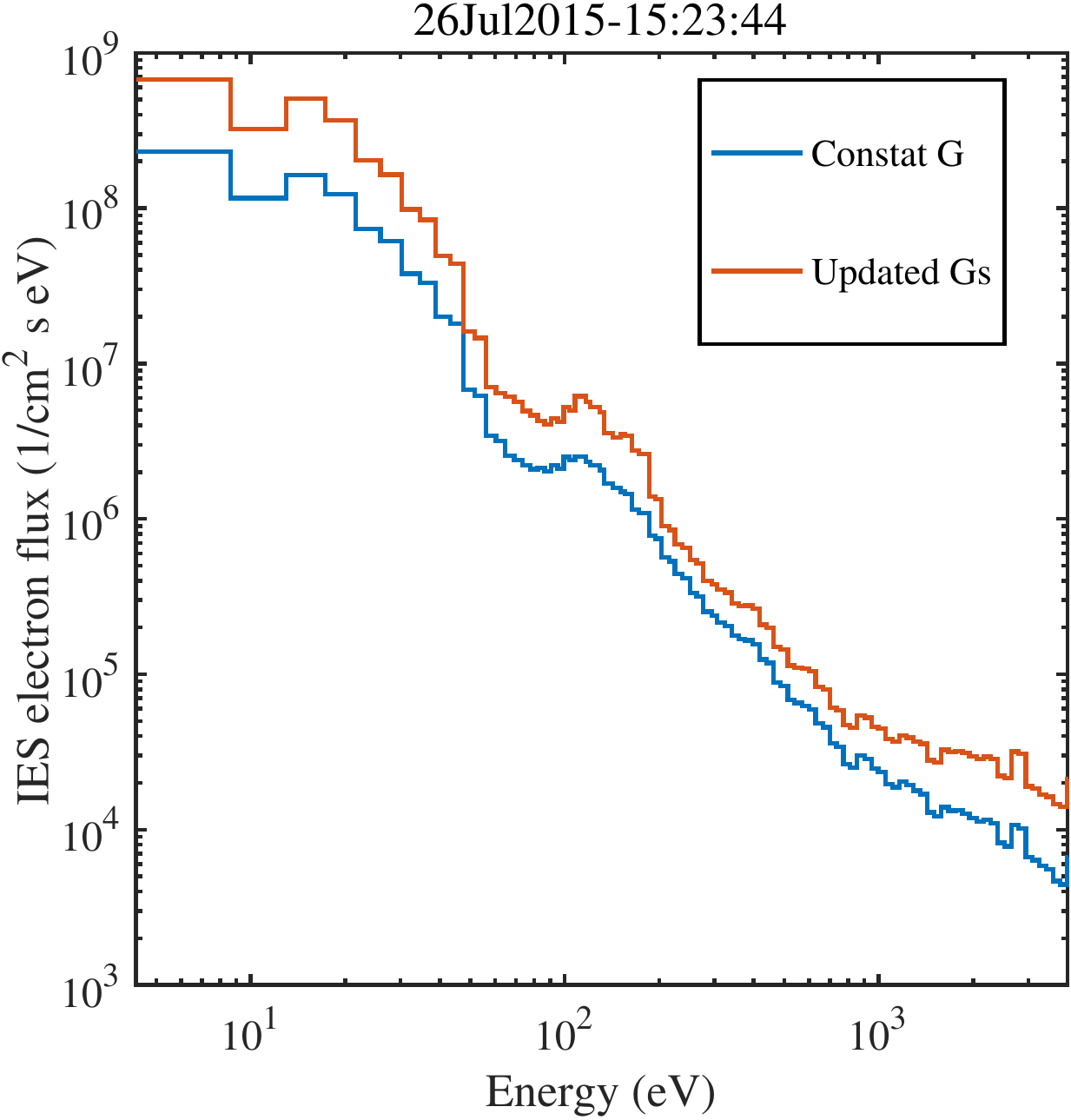}
\caption{Comparison of IES differential flux energy spectra using the updated geometric factors (red) and the initial geometric factor of the instrument paper (blue). Fluxes are summed over the entire FOV.}
\label{figGcompare_totflx}
\end{figure}

\end{document}